# On the covariant representation of integral equations of the electromagnetic field

Sergey G. Fedosin

PO box 614088, Sviazeva str. 22-79, Perm, Perm Krai, Russia

E-mail: fedosin@hotmail.com

Gauss integral theorems for electric and magnetic fields, Faraday's law of electromagnetic induction, magnetic field circulation theorem, theorems on the flux and circulation of vector potential, which are valid in curved spacetime, are presented in a covariant form. Covariant formulas for magnetic and electric fluxes, for electromotive force and circulation of the vector potential are provided. In particular, the electromotive force is expressed by a line integral over a closed curve, while in the integral, in addition to the vortex electric field strength, a determinant of the metric tensor also appears. Similarly, the magnetic flux is expressed by a surface integral from the product of magnetic field induction by the determinant of the metric tensor. A new physical quantity is introduced – the integral scalar potential, the rate of change of which over time determines the flux of vector potential through a closed surface. It is shown that the commonly used four-dimensional Kelvin-Stokes theorem does not allow one to deduce fully the integral laws of the electromagnetic field and in the covariant notation requires the addition of determinant of the metric tensor, besides the validity of the Kelvin-Stokes theorem is limited to the cases when determinant of metric tensor and the contour area are independent from time. This disadvantage is not present in the approach that uses the divergence theorem and equation for the dual electromagnetic field tensor. The problem of interpreting the law of electromagnetic induction and magnetic field circulation theorem cannot be solved on the basis of the Lorentz force in the absence of charges, and therefore requires a more general approach, when transformation of the field components from the reference frame at rest into the moving reference frame is taken into account. A new effect is predicted, according to which the circulation of magnetic field can appear even in the absence of electric current and with a constant electric field through the contour, if the area of this contour would change. By analogy with electromagnetic induction, for the magnetic field circulation to appear it is important that electric field flux that passes through the area of the contour would change over time.

**Keywords:** Gauss integral theorem; electromagnetic induction; circulation theorem; Kelvin-Stokes theorem; divergence theorem; magnetic flux; electric flux; electromotive force; vector potential; dual electromagnetic tensor.

## 1. Introduction

In classical electrodynamics, the electromagnetic field equations are written in the form of Maxwell equations:

$$\nabla \cdot \mathbf{E} = \frac{\rho}{\varepsilon_0}, \tag{1}$$

$$\nabla \cdot \mathbf{B} = 0, \tag{2}$$

$$\nabla \times \mathbf{E} = -\frac{\partial \mathbf{B}}{\partial t}, \tag{3}$$

$$\nabla \times \mathbf{B} = \mu_0 \mathbf{j} + \frac{1}{c^2} \frac{\partial \mathbf{E}}{\partial t}, \tag{4}$$

where $\mathbf{E}$ and $\mathbf{B}$ represent three-dimensional vectors of the electric field strength and magnetic field induction, respectively; $\rho$ is the charge density of a moving matter element from the viewpoint of a fixed observer; $\mathbf{j}$ denotes the density of the electric current; $\varepsilon_0$ is the electric constant; $\mu_0$ is the magnetic constant; $c$ is the speed of light, while $\mu_0 \varepsilon_0 c^2 = 1$.

The divergence theorem relates the integral of the divergence of a vector over the arbitrary three-dimensional volume $V$ to the flux of the given vector through the total surface $S$ of the given volume. In particular, for the electric field we can write the following:

$$\int_V \nabla \cdot \mathbf{E} \, dV = \oint_S \mathbf{E} \cdot \mathbf{n} \, dS = \oint_S \left( E_x \, dydz + E_y \, dzdx + E_z \, dxdy \right), \tag{5}$$

where $\mathbf{n}$ is an outward-directed unit normal vector to the surface $S$, and in Cartesian coordinates the expression $\mathbf{E} = (E_x, E_y, E_z)$ defines the components of the electric field strength vector.

If we integrate (1) over a certain volume, inside which the charge $q$ is placed, and apply (5), we will obtain an integral equation in the form of the Gauss theorem:

$$\oint_S \mathbf{E} \cdot \mathbf{n}\, dS = \frac{1}{\varepsilon_0} \int_V \rho\, dV = \frac{q}{\varepsilon_0}. \tag{6}$$

In the general case, the charges are moving, and therefore $dV$ in the integral in (6) is a volume element of a moving matter element. Due to the Lorentz contraction, both the charge density $\rho$ and $dV$ change, however, the combination $\rho\, dV$ after integration leads to the invariant charge value $q$, and therefore to conservation of the flux $\mathbf{E}$ through the surface surrounding the given volume.

If in (5) we replace $\mathbf{E}$ by $\mathbf{B}$ and take into account (2), we obtain the Gauss theorem for the magnetic field:

$$\oint_S \mathbf{B} \cdot \mathbf{n}\, dS = 0. \tag{7}$$

Integral equation (7) indicates the general absence of magnetic charges in any volume, which, if they existed, could create a non-zero magnetic induction flux through a closed surface $S$.

When integrating equation (3), the Kelvin-Stokes theorem is used, which relates the flux of the curl of the vector field through a certain two-dimensional surface $S$ to the circulation of this field around the one-dimensional contour $\ell$ bounding the given surface. For the electric field we obtain the following:

$$\int_S (\nabla \times \mathbf{E}) \cdot \mathbf{n}\, dS = \int_S \left[ (\nabla \times \mathbf{E})_x\, dydz + (\nabla \times \mathbf{E})_y\, dzdx + (\nabla \times \mathbf{E})_z\, dxdy \right] = \oint_\ell \mathbf{E} \cdot d\mathbf{r}. \tag{8}$$

The Faraday's law of electromagnetic induction follows from (3) and (8):

$$\oint_\ell \mathbf{E} \cdot d\mathbf{r} = \varepsilon_F = -\int_S \frac{\partial \mathbf{B}}{\partial t} \cdot \mathbf{n}\, dS. \tag{9}$$

As a rule, the surface $S$ and the contour $\ell$ in the Kelvin-Stokes theorem are considered to be stationary, so that, according to (9), the vortex electric field $\mathbf{E}$ appears in the contour due to changing over time of the magnetic field $\mathbf{B}$ passing through the surface. When there are charges that can move under the action of this field $\mathbf{E}$, this suggests the existence of the electromotive force $\varepsilon_F$ in the selected contour.

If we substitute $\mathbf{B}$ instead of $\mathbf{E}$ into (8) and use the result in order to integrate (4) over the surface, we will obtain the expression of the integral theorem on the circulation of the magnetic field around the contour $\ell$:

$$\oint_\ell \mathbf{B} \cdot d\mathbf{r} = \mu_0 \int_S \mathbf{j} \cdot \mathbf{n} \, dS + \frac{1}{c^2} \int_S \frac{\partial \mathbf{E}}{\partial t} \cdot \mathbf{n} \, dS = \mu_0 I_\perp + \frac{1}{c^2} \int_S \frac{\partial \mathbf{E}}{\partial t} \cdot \mathbf{n} \, dS. \qquad (10)$$

In (10) $I_\perp = \int_S \mathbf{j} \cdot \mathbf{n} \, dS = \int_S j_\perp \, dS$ denotes the total perpendicular component of the strength of current passing through the surface $S$, $\frac{\partial \mathbf{E}}{\partial t}$ defines the rate of change over time of the electric field crossing the surface. Both $I_\perp$ and the flux of $\frac{\partial \mathbf{E}}{\partial t}$ generate the vortex magnetic field $\mathbf{B}$ in the contour $\ell$ bounding the surface $S$.

The above standard formulas (1-10) are written for three-dimensional vectors. We mentioned them here so that we can then compare them with the formulas for four-dimensional vectors and tensors, which are valid not only in the flat Minkowski spacetime, but also in the curved spacetime. As will be shown later, obtaining the four-dimensional integral equations of the electromagnetic field allows us not only to generalize the three-dimensional integral equations, but also to obtain a more accurate description of the phenomena.

In particular, we will show how the problem of interpreting the law of electromagnetic induction can be solved [1]. Indeed, according to (9), the electromotive force $\varepsilon_F$ in the fixed contour arises due to changing over time of the magnetic field through the area inside the contour. However, it is well known and proved by experiments that change in the contour's area with a constant magnetic field also creates the electromotive force in the contour. Thus, in (9), the time derivative should be not inside, but before the integral over the area of the contour. Since this is not taken into account in (9), a different explanation is usually used, involving the Lorentz force [2]. At the same time, the advantage of the four-dimensional description is the possibility to take into account in one formula both sides of the electromagnetic induction, including changes in the magnetic field and in the contour's area.

The peculiar feature of the presented approach is that the main consequences arise after applying the divergence theorem to the four-dimensional electromagnetic field equations containing the electromagnetic field tensor $F_{\mu\nu}$ and its dual tensor $\tilde{F}^{\alpha\beta}$. In this case, it becomes clear that the four-dimensional Kelvin-Stokes theorem can be obtained by simplifying the divergence theorem, and therefore it is not required to derive the four-dimensional integral equations of the electromagnetic field.

In the last section, we will analyze the integral theorem on the magnetic field circulation and a conclusion will be made that with an increase in the contour's area in the constant electric field, the magnetic field and its circulation must arise in this contour.

Differential and integral equations of the electromagnetic field in the four-dimensional curved spacetime have been repeatedly considered before. If we take recent works, then, for example, in [3-5] all the variables and equations for them have been presented by splitting the components in the form 3+1 from the viewpoint of the local observer's reference frame, next to which the considered volume element with the matter and field is moving. In [6] quantum equations of the electromagnetic field are considered.

In contrast, our task is to present all the electromagnetic quantities and their integrals in a covariant vector-tensor form in an arbitrary reference frame for macroscopic systems, without first splitting vectors and tensors into separate components.

## 2. The covariant electromagnetic field equations

The four-dimensional formulation takes into account the fact that the components of the electromagnetic tensor $F_{\mu\nu}$ depend only on the components of the vectors $\mathbf{E}$ and $\mathbf{B}$. As for the charge density $\rho$ and the electric current density $\mathbf{j}$, they are included in the four-current $j^\alpha = (c\rho, \mathbf{j}) = \rho_0 u^\alpha$, where $\rho_0$ is the charge density in the comoving reference frame of the matter element, $u^\alpha$ is the four-velocity of the matter element. As a result, equations (1) and (4) are replaced by one equation:

$$\nabla_\beta F^{\alpha\beta} = -\mu_0 j^\alpha. \qquad (11)$$

In turn, equations (2) and (3) can be derived from the equation:

$$\nabla_\alpha F_{\beta\gamma} + \nabla_\beta F_{\gamma\alpha} + \nabla_\gamma F_{\alpha\beta} = \partial_\alpha F_{\beta\gamma} + \partial_\beta F_{\gamma\alpha} + \partial_\gamma F_{\alpha\beta} = 0. \qquad (12)$$

For this in (12) we should use not coinciding with each other values of the indices $\alpha, \beta, \gamma$. Another method involves the use for the pseudo-Euclidean spacetime of the completely antisymmetric Levi-Civita symbol $\varepsilon^{\alpha\beta\chi\delta}$, which implies by definition $\varepsilon^{0123} = \sqrt{-g}$, where $g$ is the determinant of the metric tensor $g_{\alpha\beta}$. If the dual electromagnetic field tensor is given by the expression

$$\tilde{F}^{\alpha\beta} = \frac{1}{2}\varepsilon^{\alpha\beta\gamma\delta}F_{\gamma\delta}, \tag{13}$$

and if we multiply all the terms in (12) by $\frac{1}{2}\varepsilon^{\alpha\delta\beta\gamma}$ and sum them up, we will obtain $-3\nabla_\alpha \tilde{F}^{\delta\alpha} = 0$. Thus, we can assume that equations (2) and (3) follow from the equation:

$$\nabla_\beta \tilde{F}^{\alpha\beta} = 0, \tag{14}$$

where the dual tensor, according to (13), has the following components:

$$\tilde{F}^{\alpha\beta} = \sqrt{-g}\begin{pmatrix} 0 & -B_x & -B_y & -B_z \\ B_x & 0 & \dfrac{E_z}{c} & -\dfrac{E_y}{c} \\ B_y & -\dfrac{E_z}{c} & 0 & \dfrac{E_x}{c} \\ B_z & \dfrac{E_y}{c} & -\dfrac{E_x}{c} & 0 \end{pmatrix}. \tag{15}$$

Two four-dimensional equations (11) and (14) replace the four three-dimensional Maxwell equations and represent the standard way to write the electromagnetic field equations in the curved spacetime.

Since the electromagnetic field tensor is an antisymmetric tensor, its covariant derivative can be represented in terms of the partial derivative:

$$\nabla_\beta F^{\alpha\beta} = \frac{1}{\sqrt{-g}}\partial_\beta\left(\sqrt{-g}\, F^{\alpha\beta}\right). \tag{16}$$

Let us multiply both sides of (16) by the covariant element of the four-volume $d\Sigma = \sqrt{-g}\, dx^0 dx^1 dx^2 dx^3$, take the volume integral and use the divergence theorem in the four-dimensional form:

$$\int_\Sigma \nabla_\beta F^{\alpha\beta} d\Sigma = \int_\Sigma \partial_\beta \left(\sqrt{-g}\, F^{\alpha\beta}\right) dx^0 dx^1 dx^2 dx^3 = \int_{S_\beta} F^{\alpha\beta} \sqrt{-g}\, dS_\beta, \qquad (17)$$

where $dS_\beta = n_\beta dS$ is the orthonormal differential $dS$ of the three-dimensional hypersurface, surrounding the physical system in the four-dimensional space, $n_\beta$ is the four-dimensional normal vector, perpendicular to the hypersurface and outward-directed.

Now we will multiply the right-hand side of (11) by $d\Sigma$ and take the integral over the four-volume:

$$-\mu_0 \int_\Sigma j^\alpha d\Sigma = -\mu_0 \int \left[\int_V j^\alpha \sqrt{-g}\, dx^1 dx^2 dx^3 \right] dx^0. \qquad (18)$$

Comparison of (11), (17) and (18) gives the following:

$$\int_{S_\beta} F^{\alpha\beta} \sqrt{-g}\, dS_\beta = \int_V F^{\alpha 0} \sqrt{-g}\, dx^1 dx^2 dx^3 + \int \left[\int F^{\alpha 1} \sqrt{-g}\, dx^2 dx^3 \right] dx^0 +$$

$$+ \int \left[\int F^{\alpha 2} \sqrt{-g}\, dx^3 dx^1 \right] dx^0 + \int \left[\int F^{\alpha 3} \sqrt{-g}\, dx^1 dx^2 \right] dx^0 = -\mu_0 \int \left[\int_V j^\alpha \sqrt{-g}\, dx^1 dx^2 dx^3 \right] dx^0.$$

Let us differentiate this equality with respect to the variable $x^0 = ct$, where $t$ is the coordinate time:

$$\frac{1}{c}\frac{d}{dt}\int_V F^{\alpha 0} \sqrt{-g}\, dx^1 dx^2 dx^3 + \int F^{\alpha 1} \sqrt{-g}\, dx^2 dx^3 + \int F^{\alpha 2} \sqrt{-g}\, dx^3 dx^1 + \int F^{\alpha 3} \sqrt{-g}\, dx^1 dx^2 =$$

$$= -\mu_0 \int_V j^\alpha \sqrt{-g}\, dx^1 dx^2 dx^3.$$

(19)

At $\alpha = 0$ the first term in (19) vanishes, since $F^{00} = 0$. For the remaining terms, in view of the equality $j^0 = \rho_0 u^0$, we can write the following:

$$\oint_{S_k} F^{0k} \sqrt{-g}\, dS_k = -\frac{\Phi_E}{c} = -\mu_0 \int_V \rho_0 u^0 \sqrt{-g}\, dx^1 dx^2 dx^3 = -c\mu_0 \int_{V_0} \rho_0 dV_0 = -c\mu_0 q. \qquad (20)$$

Here $dS_k = dS^{ij} = dx^i dx^j$ is an orthonormal element of the two-dimensional surface, surrounding the charge $q$; $\Phi_E = c \oint_{S_k} F^{k0} \sqrt{-g}\, dS_k$ represents the electromagnetic field flux through the closed surface; the three-dimensional indices $i, j, k = 1, 2, 3$ and they do not coincide with each other.

The relation from [7] was also used for the volume element $dV_0$ in the comoving reference frame of this element:

$$\frac{dt}{d\tau} \sqrt{-g}\, dx^1 dx^2 dx^3 = \frac{u^0}{c} \sqrt{-g}\, dx^1 dx^2 dx^3 = dV_0. \qquad (21)$$

Integral equation (20) is the Gauss theorem in the covariant notation and generalizes equation (6).

Let us now assume that in (19) $\alpha = k = 1, 2, 3$:

$$\frac{1}{c}\frac{d}{dt} \int_V F^{k0} \sqrt{-g}\, dx^1 dx^2 dx^3 + \int F^{k1} \sqrt{-g}\, dx^2 dx^3 + \int F^{k2} \sqrt{-g}\, dx^3 dx^1 + \int F^{k3} \sqrt{-g}\, dx^1 dx^2 =$$
$$= -\mu_0 \int_V j^k \sqrt{-g}\, dx^1 dx^2 dx^3.$$

$$(22)$$

Let us consider (22) at $k = 1$. Since $F^{\alpha\beta} = g^{\alpha\gamma} g^{\beta\delta} F_{\gamma\delta}$ by definition, then at $\alpha = 1$, $\beta = 1$ it will be $F^{11} = g^{1\gamma} g^{1\delta} F_{\gamma\delta}$. Then, due to symmetry of the metric tensor $g^{\alpha\gamma}$ and antisymmetry of the electromagnetic field tensor $F_{\gamma\delta}$, it turns out that in (22) $F^{11} = 0$.

Suppose now that the volume under consideration is such that its size in the direction of the axis $OX$ equals $a$ and is small in value, so that $a \approx 0$. Then the result of integration over the differential $dx^1 = dx$ in (22) can be considered as the product of the integrands by $a$. Since $F^{11} = 0$, then in (22) it will be possible to reduce all the terms by $a$. First we will replace the area element $dx^1 dx^2$ by $-dx^2 dx^1$, which will change the sign in the corresponding surface integral of the second kind. The following remains:

$$\frac{1}{c}\frac{d}{dt}\int_S F^{10}\sqrt{-g}\,dx^2 dx^3 + \int F^{12}\sqrt{-g}\,dx^3 - \int F^{13}\sqrt{-g}\,dx^2 = -\mu_0 \int_S j^1 \sqrt{-g}\,dx^2 dx^3 \,.$$

The quantity $\Phi_{Ex} = c\int_S F^{10}\sqrt{-g}\,dx^2 dx^3$ here can be considered as the electric field flux through the surface in the direction of the axis $OX$, since the main contribution to $F^{10}$ is made by the component $F_{01}$, which is associated with the electric field component $E_x$. In the general case, each of the components $F^{10}$, $F^{12}$ and $F^{13}$ depends on all the components of the tensor $F_{\gamma\delta}$ at the same time. The situation can be simplified within the framework of the special theory of relativity, where $g^{\alpha\gamma} = \eta^{\alpha\gamma}$, in which case the components of the metric tensor $\eta^{\alpha\gamma}$ in the Minkowski spacetime are equal to 0 or 1 and do not depend on the time and coordinates. In this case, we will obtain $\sqrt{-g}=1$, $F^{10}=\dfrac{E_x}{c}$, $F^{12}=-B_z$, $F^{13}=B_y$, $j^1=j_x$, and the integral equation becomes as follows:

$$\frac{1}{c^2}\frac{d}{dt}\int_S E_x\,dy\,dz - \int B_z\,dz - \int B_y\,dy = -\mu_0 \int_S j_x\,dy\,dz = -\mu_0 I_x \,.$$

This can be rewritten similarly to (10):

$$\int B_z\,dz + \int B_y\,dy = \oint_\ell \mathbf{B}\cdot d\mathbf{r} = \mu_0 I_x + \frac{1}{c^2}\frac{d}{dt}\int_S E_x\,dy\,dz \,. \tag{23}$$

In (23) the contour $\ell$ and the surface $S$ are located in the plane $YOZ$, and the magnetic field circulation around this contour arises due to the current $I_x$ through the area $S$ inside the contour, as well as in case when there is change over time in the electric field flux with the strength $E_x$ crossing the area $S$.

Integral equation (10) was derived using the Kelvin-Stokes theorem, and equations (22) and (23) follow from the divergence theorem. This implies close relation between these theorems, since it can be seen that for the Kelvin-Stokes theorem to hold true, it is necessary that the thickness of the volume under consideration should tend to zero everywhere, regardless of the orientation of the parts of this volume in space. However, there is a difference between (23) and (10), which consists in the fact that in (23) the time derivative of the total electric field flux over the surface is taken, and in (10), instead, there is only the partial time derivative of the electric field with the constant surface area. As a result, the integral theorem on the magnetic field circulation in the form of (22) is more informative than (10), since (22) includes integrals over the volume and the total time derivative, and (10) includes only integrals over the surface.

### 3. The Kelvin-Stokes theorem in the four-dimensional form

The Kelvin-Stokes theorem (8) relates the integral of the flux of the curl of the three-dimensional vector over a certain area to the circulation of this vector around the contour bounding the specified area. The four-dimensional generalization of theorem (8) can be found, for example, in [8]:

$$\frac{1}{2}\int_S \left(\nabla_\alpha \times A_\beta\right) dS^{\alpha\beta} = \oint_\ell A_\beta dx^\beta. \tag{24}$$

The area integral in (24) contains the four-curl of the four-vector $A_\beta$:

$$\nabla_\alpha \times A_\beta = \nabla_\alpha A_\beta - \nabla_\beta A_\alpha = \partial_\alpha A_\beta - \partial_\beta A_\alpha.$$

However, as we will show below, (24) is valid only within the framework of the special theory of relativity, and in the curved spacetime it is necessary to introduce in (24) the determinant $g$ of the metric tensor:

$$\frac{1}{2}\int_S g\left(\nabla_\alpha \times A_\beta\right) dS^{\alpha\beta} = \oint_\ell g A_\beta dx^\beta. \tag{25}$$

It is assumed that the area element in (25) is antisymmetrically oriented, so that the following relations hold true:

$$dS^{01} = dt\,dx, \qquad dS^{02} = dt\,dy, \qquad dS^{03} = dt\,dz, \qquad dS^{12} = dx\,dy,$$

$$dS^{13} = -dx\,dz, \qquad dS^{23} = dy\,dz, \qquad dS^{\alpha\beta} = -dS^{\beta\alpha}. \tag{26}$$

We will consider the case when in (25) the indices $\alpha, \beta = 1, 2, 3$, and the four-vector $A_\beta$ is the four-potential of the electromagnetic field, considered at a certain time point. Then $A_\beta = \left(\dfrac{\varphi}{c}, -\mathbf{A}\right)$, where $\varphi$ is the scalar potential, $\mathbf{A}$ is the vector potential. In this case $\nabla_\alpha \times A_\beta = F_{\alpha\beta}$, and from (25) and (26) it follows:

$$\int_S g\left(B_z\, dxdy + B_y\, dzdx + B_x\, dydz\right) = \int_S g\mathbf{B}\cdot\mathbf{n}\, dS = -\Phi = \int_S g(\nabla\times\mathbf{A})\cdot\mathbf{n}\, dS = \oint_\ell g\left(A_x\, dx + A_y\, dy + A_z\, dz\right) = \oint_\ell g\mathbf{A}\cdot d\mathbf{r}. \tag{27}$$

In (27) it was taken into account that $\mathbf{B} = \nabla\times\mathbf{A}$. According to (27), the magnetic field flux $\Phi = -\int_S g\mathbf{B}\cdot\mathbf{n}\, dS$ through a certain fixed surface $S$ leads to circulation of the vector potential $\mathbf{A}$ around a fixed contour, surrounding this surface.

Let us take the partial derivative with respect to time in (27) and rearrange the terms:

$$\oint_\ell \frac{\partial(g\mathbf{A})}{\partial t}\cdot d\mathbf{r} = \oint_\ell \mathbf{E}_A\cdot d\mathbf{r} = \varepsilon_A = \int_S \frac{\partial(g\mathbf{B})}{\partial t}\cdot\mathbf{n}\, dS. \tag{28}$$

The index $A$ in (28) shows that the circular electric field $\mathbf{E}_A$ and the electromotive force $\varepsilon_A$ in the contour $\ell$ arise due to changing over time of the magnetic field, passing through the contour, which generates a change in the circulation of the vector potential $\mathbf{A}$. Since, by definition, the electric field appears in the presence of the gradient of the scalar potential and change over time of the vector potential, $\mathbf{E} = -\nabla\varphi - \dfrac{\partial\mathbf{A}}{\partial t}$, then the vortex field $\mathbf{E}_A = \dfrac{\partial(g\mathbf{A})}{\partial t}$ and the electromotive force $\varepsilon_A$ in (28) appear in the absence of the scalar potential $\varphi$. From comparison of (28) and (9) now we can see that the electromagnetic induction law can be represented as a special case of (25).

Let us write (27) for a practically closed surface, when the contour $\ell$ becomes so small that its length can be considered equal to zero:

$$\oint_S g\left(B_z\, dxdy + B_y\, dxdz + B_x\, dydz\right) = \oint_S g\mathbf{B}\cdot\mathbf{n}\, dS = -\Phi = \oint_\ell g\mathbf{A}\cdot d\mathbf{r} = 0. \tag{29}$$

According to (29), the magnetic field flux $\Phi$ through a closed surface is equal to zero. In the limit of the special theory of relativity $g=-1$ and integral equation (29) turns into the Gauss theorem for the magnetic field in (7).

Let us now consider the case when in (25), in view of the relation $\nabla_\alpha \times A_\beta = F_{\alpha\beta}$ and (26-27), the indices range over all the values $\alpha, \beta = 0, 1, 2, 3$:

$$\int g\left(E_x\, dt\,dx + E_y\, dt\,dy + E_z\, dt\,dz - B_z\, dx\,dy - B_y\, dz\,dx - B_x\, dy\,dz\right) = \int \left(\oint_\ell g\mathbf{E}\cdot d\mathbf{r}\right) dt -$$

$$-\int_S g(\nabla \times \mathbf{A})\cdot(\mathbf{n}\,dS) = -\int \varepsilon\, dt - \oint_\ell g\left(A_x\, dx + A_y\, dy + A_z\, dz\right) = \int g\varphi\, dt - \oint_\ell g\mathbf{A}\cdot d\mathbf{r}.$$

(30)

In (30) the electromotive force $\varepsilon = -\oint_\ell g\mathbf{E}\cdot d\mathbf{r}$ was introduced. For integral equation (30) to hold, it is necessary that $\int \varepsilon\, dt = -\int g\varphi\, dt$, that is, the electromotive force behaves as a potential with accuracy up to the multiplier $-g$: $\varepsilon = -g\varphi$. It should be noted that in order for the electromotive forces $\varepsilon$ and $\varepsilon_A$ from (28) to coincide, it is necessary that the relation $\mathbf{E}_A = \dfrac{\partial(g\mathbf{A})}{\partial t} = -g\mathbf{E}$ should hold, which is possible if the determinant $g$ does not depend on the time, and the vortex field $\mathbf{E}$ in (30) is generated only by change over time of the vector potential $\mathbf{A}$.

### 4. Integral equations for the dual electromagnetic tensor

Let us apply the divergence theorem to the dual tensor $\tilde{F}^{\alpha\beta}$ in (14), and acting similarly to (16) and (17), we find:

$$\int_\Sigma \nabla_\beta \tilde{F}^{\alpha\beta}\, d\Sigma = \int_\Sigma \partial_\beta \left(\sqrt{-g}\, \tilde{F}^{\alpha\beta}\right) dx^0\, dx^1\, dx^2\, dx^3 = \int_{S_\beta} \tilde{F}^{\alpha\beta} \sqrt{-g}\, dS_\beta = 0. \qquad (31)$$

We will write this integral equation by the components:

$$\int_V \tilde{F}^{\alpha 0} \sqrt{-g}\, dx^1\, dx^2\, dx^3 + \int\left[\int \tilde{F}^{\alpha 1} \sqrt{-g}\, dx^2\, dx^3\right] dx^0 +$$

$$+ \int\left[\int \tilde{F}^{\alpha 2} \sqrt{-g}\, dx^3\, dx^1\right] dx^0 + \int\left[\int \tilde{F}^{\alpha 3} \sqrt{-g}\, dx^1\, dx^2\right] dx^0 = 0,$$

and then will take the derivative with respect to the variable $x^0 = ct$:

$$\frac{1}{c}\frac{d}{dt}\int_V \tilde{F}^{\alpha 0}\sqrt{-g}\,dx^1 dx^2 dx^3 + \int \tilde{F}^{\alpha 1}\sqrt{-g}\,dx^2 dx^3 + \int \tilde{F}^{\alpha 2}\sqrt{-g}\,dx^3 dx^1 + \int \tilde{F}^{\alpha 3}\sqrt{-g}\,dx^1 dx^2 = 0.$$

(32)

Equation (32) with the index $\alpha = 0$, in view of the dual tensor components in (15), can be written as follows:

$$-\oint_S g\left(B_x\,dydz + B_y\,dzdx + B_z\,dxdy\right) = -\oint_S g\,\mathbf{B}\cdot\mathbf{n}\,dS = \Phi = 0, \tag{33}$$

where $g$ is the determinant of the metric tensor $g_{\alpha\beta}$.

Integral equation (33), just as (29), represents the Gauss theorem for the magnetic field, and, in the limit of the special theory of relativity, it turns into (7).

Let us now assume that in (32) the index $\alpha = 1$ and use (15):

$$\frac{d}{dt}\int_V B_x\,g\,dxdydz + \int E_z\,g\,dzdx - \int E_y\,g\,dxdy = 0. \tag{34}$$

Suppose that the size $a$ of the volume under consideration in the direction of the axis $OX$ is so small that the integrals in (34) can be represented as the products of the integrands by $a$. Then, taking into account the equality $\int E_y\,g\,dxdy = -\int E_y\,g\,dydx$, all the terms in (34) can be reduced by $a$ and the following remains:

$$\frac{d}{dt}\int_S B_x\,g\,dydz + \int E_z\,g\,dz + \int E_y\,g\,dy = 0.$$

This can be rewritten in a vector form, given that $d\mathbf{r} = (dx, dy, dz)$:

$$-\oint_\ell g\mathbf{E}\cdot d\mathbf{r} = \varepsilon = \frac{d}{dt}\int_S g\mathbf{B}\cdot\mathbf{n}\,dS = -\frac{d\Phi}{dt}, \qquad (35)$$

For the above case, when in (32) the index $\alpha = 1$ and the contour $\ell$ is located in the plane $YOZ$, if the contour is circuited in a counterclockwise manner, then $d\mathbf{r} = (0, dy, dz)$, and the vector $\mathbf{n}$ is directed along the axis $OX$. In this case, direction of the field $\mathbf{E}$ would coincide with the counterclockwise direction of circulation around the contour, if the field $\mathbf{B}$ would increase with time and would be directed against the axis $OX$.

Integral equation (35) represents the Faraday's law of electromagnetic induction in a covariant form, where the quantity $\Phi = -\int_S g\mathbf{B}\cdot\mathbf{n}\,dS$ is the magnetic flux through the surface $S$ bounded by the conducting contour $\ell$.

It should be noted that the magnetic flux in (35) can change not only when the magnetic field $\mathbf{B}$ changes with time, but also when the area of the surface $S$ changes. Thus, changes of both $\mathbf{B}$ and $S$ can contribute to the circulation of the electric field around the contour and to creation of the electromotive force $\varepsilon$. By contrast, in integral equations (9) and (28), obtained from the three-dimensional and four-dimensional Kelvin-Stokes theorems, respectively, the electromotive force arises only due to changing over time of the vector potential of the magnetic field in the fixed contour of constant area.

### 5. Analysis of the integral theorem on the electric field circulation

In the previous section, we pointed to the limited use of the Kelvin-Stokes theorem to describe the effect of electromagnetic induction. Complete description of this effect can be achieved by applying the divergence theorem to the dual electromagnetic field tensor, which leads to integral equation (35). We will apply (35) to describe the standard experiment with a frame, in which there is a moving crossbar, allowing us to change the area of the frame. The frame configuration is shown in Figure 1.

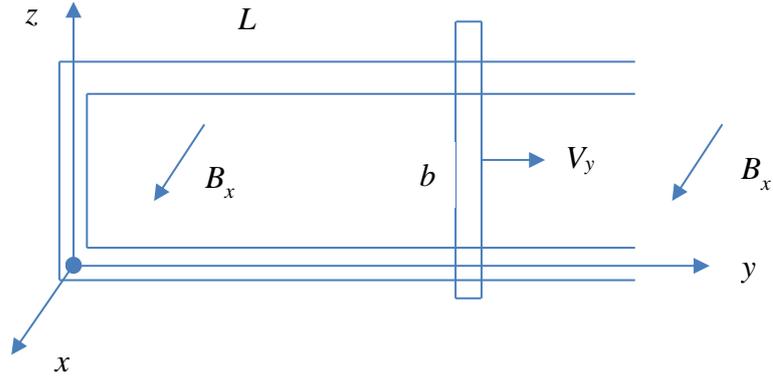

Fig. 1. A frame with a moving crossbar, the size of which is equal to $b$.

Let the height of the crossbar be equal to $b$, and the current distance from the origin of coordinates to the crossbar be equal to $L = V_y t$, where $V_y$ is the velocity of the crossbar's motion along the axis $OY$. The constant magnetic field $B_x$ is directed along the axis $OX$. Within the framework of the special theory of relativity, the determinant of the metric tensor $g = -1$, and the magnetic flux is calculated as the integral of the magnetic field over the frame's area:

$$\Phi = -\int_S g \mathbf{B} \cdot \mathbf{n}\, dS = -B_x b L = -V_y B_x b t. \tag{36}$$

Here we assumed that the direction of circulation around the frame in Figure 1 is clockwise, so that, according to the right-hand screw rule, the normal $\mathbf{n}$ to the frame is directed opposite to the axis $OX$ and opposite to the vector $\mathbf{B}$. In this case, the direction of circulation should coincide with the direction of the induction current in the crossbar.

According to (35), we now find the electromotive force as the electric field circulation around the entire length of the moving crossbar:

$$-\oint_\ell g \mathbf{E} \cdot d\mathbf{r} = \int_b \mathbf{E}_b \cdot d\mathbf{r} = \varepsilon_b = -\frac{d\Phi}{dt} = V_y B_x b. \tag{37}$$

In the fixed part of the frame the magnetic field is constant and the electric field circulation there must be equal to zero. If we take into account (30), then we obtain the relation $\varepsilon_b = -g\,\varphi_b = \varphi_b$, that is, the electromotive force changes in the same way as the electric potential along the crossbar. In this case, the crossbar becomes the current generator for the entire frame, the energy source for which is the force moving the crossbar in the magnetic field. Indeed, if there is a conductive circuit and a current generator (for example, an electric battery), then inside the generator, as it moves from the cathode to the anode, the potential increases and then drops down along the circuit. We can assume that in Figure 1 the cathode of the induction current generator is at the top, and the anode is at the bottom of the crossbar.

In order to confirm this, it is necessary to imagine what happens in the crossbar as it moves. The standard interpretation of electromagnetic induction, when the area changes in the magnetic flux, is reduced to the Lorentz force acting in the magnetic field on the positive charge $q$ inside the crossbar:

$$\mathbf{F} = q[\mathbf{V} \times \mathbf{B}], \qquad F_z = -qV_y B_x. \qquad (38)$$

Next we find the electric field strength inside the crossbar directed against the axis $OZ$, while the field circulation coincides with (37):

$$\mathbf{E}_b = (0, 0, E_z), \qquad E_z = \frac{F_z}{q} = -V_y B_x, \qquad \int_b \mathbf{E}_b \cdot d\mathbf{r} = \varepsilon_b = V_y B_x b. \qquad (39)$$

The electromotive force $\varepsilon_b$ is positive, as the vector $d\mathbf{r}$ in (39) is directed inside the crossbar against the axis $OZ$, passing around the frame in a clockwise manner after the vector $\mathbf{E}_b$. In this case, the direction of circulation around the frame coincides with the direction, which was chosen in (36) to determine the direction of the normal $\mathbf{n}$ to the frame.

The situation under consideration corresponds to the Lenz rule, according to which the electromotive force arising due to induction generates in the conductive circuit such a current that its action is opposite to the action that caused the induction current. The induction effect in this case arises due to the force moving the crossbar at a constant velocity in the magnetic field, increasing the frame's area. If in (38) instead of velocity $\mathbf{V}$ we substitute the velocity $\mathbf{v}$ of the positive charge as it moves in the crossbar under the action of the strength $\mathbf{E}_b$, then we will obtain the Lorentz force $F_y = -qv_z B_x$ directed against the axis $OY$ and opposite to the force moving the crossbar.

If the crossbar moved by inertia in the absence of external forces, then the force $F_y$ in the conductive frame would lead to deceleration of the crossbar's motion and to decrease of its kinetic energy. This is due to the fact that the kinetic energy would be transformed into the energy of the induction current, released in the form of heating the conductive circuit. If the electric circuit is broken, there will be no the current in the frame, the electric field $\mathbf{E}_b$ will create in the crossbar a difference of potentials, equal to the electromotive force $\varepsilon_b$, and the crossbar will be able to move at a certain constant velocity even in the absence of external forces.

In the general case, the contour or frame under consideration may be non-conductive and there may be no both free charges and bound charges in them. Then the explanation of the induction effect based on the Lorentz force according to (38-39) becomes inapplicable, although the electric field circulation and the corresponding electromotive force always take place in the moving crossbar. Consequently, a different, more general explanation is required, for example, based on transformation of the components of the electric and magnetic fields from the reference frame $K$, associated with the fixed frame, into the reference frame $K'$, associated with the moving crossbar.

The electric field $\mathbf{E}$ and magnetic field $\mathbf{B}$ are part of the components of the electromagnetic field tensor, and therefore they are transformed according to the tensor law. When an arbitrary point on the crossbar moves along the axis $OY$ in Figure 1, the fields' components at this point within the framework of the special theory of relativity are transformed as follows:

$$E'_y = E_y, \qquad E'_x = \gamma E_x + \gamma V_y B_z, \qquad E'_z = \gamma E_z - \gamma V_y B_x,$$

$$B'_y = B_y, \qquad B'_x = \gamma B_x - \frac{\gamma V_y E_z}{c^2}, \qquad B'_z = \gamma B_z + \frac{\gamma V_y E_x}{c^2}. \qquad (40)$$

The quantity $\gamma = \dfrac{1}{\sqrt{1 - V^2/c^2}}$ is the Lorentz factor for the velocity $\mathbf{V} = (0, V_y, 0)$ of the crossbar's motion.

Since in $K$ the field $\mathbf{E} = 0$, the field $\mathbf{B} = (B_x, 0, 0)$, then in $K'$ we will have only the following non-zero field components: $E'_z = -\gamma V_y B_x$, $B'_x = \gamma B_x$. Therefore, an observer moving in $K'$ with the crossbar should see the electric field circulation in the form

$$\int_b \mathbf{E}'_b \cdot d\mathbf{r}' = \varepsilon'_b = \gamma V_y B_x b. \tag{41}$$

From comparison of (37) and (41) it follows that $\varepsilon'_b = \gamma \varepsilon_b$. Contribution to the electromotive forces $\varepsilon_b$ and $\varepsilon'_b$ is made only by the electric fields $\mathbf{E}_b$ and $\mathbf{E}'_b$, respectively. This distinguishes the electromotive force from the electric potential $\varphi$, since, according to the definition $\mathbf{E} = -\nabla \varphi - \frac{\partial \mathbf{A}}{\partial t}$, the field $\mathbf{E}$ is associated not only with $\varphi$, but also with the vector potential $\mathbf{A}$. However, in the case under consideration, the magnetic field $\mathbf{B}$ does not depend on the time and, since $\mathbf{B} = \nabla \times \mathbf{A}$, then $\mathbf{A}$ does not depend on the time either, not contributing to the electric field. That is why, according to (30), the electromotive force $\varepsilon_b$ behaves similarly to $\varphi_b$.

Let us now consider how the electric potential should be transformed from $K$ into the reference frame $K'$. If in $K$ there is the four-potential of the electric field $A_\alpha = \left( \frac{\varphi}{c}, -\mathbf{A} \right)$, and $K'$ moves in $K$ along the axis $OY$, then according to the Lorentz transformations for the four-vectors we obtain the following:

$$\varphi' = \gamma \varphi - \gamma V_y A_y, \qquad A'_x = A_x, \qquad A'_y = \gamma A_y - \frac{\gamma \varphi V_y}{c^2}, \qquad A'_z = A_z.$$

Using the freedom to choose the vector potential associated with gauging of the four-potential in the form $\partial^\alpha A_\alpha = \partial_\alpha A^\alpha = \frac{1}{c^2}\frac{\partial \varphi}{\partial t} + \nabla \cdot \mathbf{A} = 0$ in the flat Minkowski spacetime, in $K$ we can assume that $A_x = 0$, $A_y = 0$, $A_z = y B_x$, where $B_x = const$. For the components of the four-potential in $K'$ it gives:

$$\varphi' = \gamma \varphi, \qquad A'_x = 0, \qquad A'_y = -\frac{\gamma \varphi V_y}{c^2}, \qquad A'_z = y B_x.$$

The relation for the potential $\varphi' = \gamma \varphi$ is obtained in the same way as the relation $\varepsilon'_b = \gamma \varepsilon_b$ for the electromotive force obtained above. Thus, drawing an analogy between the electromotive force and the electric scalar potential, we can explain the law of electromagnetic induction for the case of increasing of the frame's area, without using the Lorentz force, but relying on the transformation of the field components between the two reference frames.

### 6. Analysis of the integral theorem on the magnetic field circulation

Since the integral theorem on the magnetic field circulation around the fixed contour $\ell$ in (10) is proved using the Kelvin-Stokes theorem, it is possible that, as is the case with the electromagnetic induction effect, the proof may not give a complete description of the phenomenon. Equation (10) states that the magnetic field circulation occurs when there is a perpendicular component of the electric current through the fixed contour, as well as in case of change over time of the electric field crossing the contour.

However, from divergence theorem (22) we obtain integral equation (23), which more fully describes the theorem on the magnetic field circulation. This equation within the framework of the special theory of relativity can be rewritten as follows:

$$\oint_\ell \mathbf{B} \cdot d\mathbf{r} = \mu_0 I_\perp + \frac{1}{c^2}\frac{d\Phi_E}{dt}, \qquad (42)$$

where $\Phi_E = \int_S \mathbf{E} \cdot \mathbf{n}\, dS$ is the flux of the electric field strength through the surface $S$ bounded by the contour $\ell$.

According to (42), the magnetic field circulation can appear in the absence of the electric current $I_\perp$ as well as with the constant electric field, if the area of the contour crossed by the electric field would change. The latter changes the flux $\Phi_E$ in (42). This conclusion cannot be predicted using the Kelvin-Stokes theorem, due to the limitations of its action.

To analyze the situation, we will refer again to Figure 1, where we will replace the magnetic field $B_x$ with the electric field $E_x$. Let us choose the counterclockwise direction of circulation around the frame, then for the left-hand side of (42) we can write:

$$\oint_\ell \mathbf{B} \cdot d\mathbf{r} = \int_b \mathbf{B}_b \cdot d\mathbf{r} = B_b b.$$

Assuming that the electric current $I_\perp$ through the frame is equal to zero, we find the right-hand side of (42):

$$\Phi_E = \int_S \mathbf{E} \cdot \mathbf{n}\, dS = E_x b L = V_y E_x b t, \qquad \frac{1}{c^2}\frac{d\Phi_E}{dt} = \frac{V_y E_x b}{c^2}.$$

As a consequence of the equality of the left-hand and right-hand sides of (42), we arrive at the relation $B_b = \dfrac{V_y E_x}{c^2}$. We cannot explain this relation by reasoning based on the Lorentz force, as was the case for the electromagnetic induction in (38). This is especially true when there are no electric charges in the selected contour. The only way to explain this is to transform the field components from $K$ into $K'$. Assuming that $\mathbf{B}=0$, $\mathbf{E}=(E_x,0,0)$ in $K$, from (40) we find in $K'$ the following non-zero field components: $E'_x = \gamma E_x$, $B'_z = \dfrac{\gamma V_y E_x}{c^2}$. Comparison with the expression for $B_b$ gives the following: $B'_z = \gamma B_b$. Note that a similar equality $E'_z = \gamma E_b$ for the case of the electromagnetic induction follows from (37) and (41).

We arrive at a certain contradiction, which consists in the fact that initially we assumed $\mathbf{B}=0$ everywhere in $K$, however, when calculating the magnetic field circulation in (42), a certain magnetic field $B_b$ appeared inside the moving crossbar. In order to avoid this contradiction, it would be more correct to assume that in fact inside the crossbar in its reference frame $K'$ there is a magnetic field $B'_z$, which creates circulation in the crossbar. While in the reference frame $K$, this circulation is manifested already as the circulation of the effective magnetic field $B_b$.

## 7. What does gauge fixing of the four-potential give us?

In the previous sections, we considered the fluxes and circulations of the electric and magnetic fields, which are the components of the electromagnetic field tensor. There is one more quantity characterizing the electromagnetic field – it is the four-potential $A_\alpha = \left(\dfrac{\varphi}{c}, -\mathbf{A}\right)$, where the components of the four-potential contain the scalar potential $\varphi$ and the vector potential $\mathbf{A}$. In (27) we determined for $\mathbf{A}$ the circulation $\oint_\ell g\mathbf{A}\cdot d\mathbf{r}$ around the fixed contour, but it turns out that it is also possible to determine the flux $\mathbf{A}$ using the closed surface.

We will start with the fact that in the covariant Lorentz gauge the four-potential must satisfy the relation:

$$\nabla^\alpha A_\alpha = \nabla_\alpha A^\alpha = \frac{1}{\sqrt{-g}} \partial_\alpha \left( \sqrt{-g}\, A^\alpha \right) = 0.$$

We will multiply this equality by the covariant element of the four-volume $d\Sigma = \sqrt{-g}\, dx^0 dx^1 dx^2 dx^3$, take the integral over the four-volume and use the divergence theorem in the four-dimensional form:

$$\int_\Sigma \nabla_\alpha A^\alpha\, d\Sigma = \int_\Sigma \partial_\alpha \left( \sqrt{-g}\, A^\alpha \right) dx^0 dx^1 dx^2 dx^3 = \int_{S_\alpha} A^\alpha \sqrt{-g}\, dS_\alpha = 0,$$

where $dS_\alpha = n_\alpha dS$ is the orthonormal differential $dS$ of the three-dimensional hypersurface, surrounding the physical system in the four-dimensional space, $n_\alpha$ is the four-dimensional normal vector, perpendicular to the hypersurface and outward-directed.

The last equality can be written in more detail:

$$\int_{S_\alpha} A^\alpha \sqrt{-g}\, dS_\alpha = \int_V A^0 \sqrt{-g}\, dx^1 dx^2 dx^3 + \int \left[ \int A^1 \sqrt{-g}\, dx^2 dx^3 \right] dx^0 +$$
$$+ \int \left[ \int A^2 \sqrt{-g}\, dx^3 dx^1 \right] dx^0 + \int \left[ \int A^3 \sqrt{-g}\, dx^1 dx^2 \right] dx^0 = 0.$$

Let us differentiate this equality with respect to the variable $x^0 = ct$, where $t$ is the coordinate time:

$$\frac{1}{c} \frac{d}{dt} \int_V A^0 \sqrt{-g}\, dx^1 dx^2 dx^3 + \int A^1 \sqrt{-g}\, dx^2 dx^3 + \int A^2 \sqrt{-g}\, dx^3 dx^1 + \int A^3 \sqrt{-g}\, dx^1 dx^2 = 0.$$

Here three area integrals in sum are equal to the integral over the closed two-dimensional surface surrounding the three-dimensional volume under consideration:

$$\oint_{S_k} A^k \sqrt{-g}\, dS_k = \Phi_A = -\frac{1}{c} \frac{d}{dt} \int_V A^0 \sqrt{-g}\, dx^1 dx^2 dx^3 = -\frac{1}{c} \frac{d}{dt} I_0, \qquad (43)$$

where $dS_k = dS^{ij}$ is an orthonormal element of the two-dimensional surface; $\Phi_A$ represents the flux of the electromagnetic field potentials through the closed surface; the indices $i, j, k = 1, 2, 3$ and they do not coincide with each other; a quantity $I_0 = \int_V A^0 \sqrt{-g}\, dx^1 dx^2 dx^3$ is the volume integral of scalar component of the four-potential and can be called the integral scalar potential.

Since $A^\alpha = g^{\alpha\beta} A_\beta$, $A^k = g^{k\beta} A_\beta$, then the components $A^k$ of the four-potential are functions of the scalar and vector potentials of the electromagnetic field.

In the limit of the special theory of relativity $\sqrt{-g} = 1$, for the Cartesian coordinates we have $A^0 = \dfrac{\varphi}{c}$, $A^k = \mathbf{A} = (A^1, A^2, A^3) = (A_x, A_y, A_z)$, and (43) is simplified:

$$\oint_S \mathbf{A} \cdot \mathbf{n}\, dS = \Phi_A = -\frac{1}{c}\frac{d}{dt} I_0 = -\frac{1}{c^2}\frac{d}{dt}\int_V \varphi\, dx^1 dx^2 dx^3 . \qquad (44)$$

In this case, we can see that $\Phi_A$ becomes the flux of the vector potential $\mathbf{A}$ over the closed surface. The flux $\Phi_A$ can appear for two reasons – either if the distribution of the scalar potential $\varphi$ inside the volume under consideration changes over time, or if the size of the volume itself changes. According to (44), the change of $\varphi$ over time can lead to appearance of the vector potential $\mathbf{A}$ in space, and if $\mathbf{A}$ would depend on time, then an additional electric field would appear due to this, according to the definition $\mathbf{E} = -\nabla \varphi - \dfrac{\partial \mathbf{A}}{\partial t}$.

## 8. Conclusion

Using the divergence theorem, we obtained four-dimensional equation (19), from which we obtained the integral Gauss theorem (20) in a covariant form. The flux of the electromagnetic field through the closed surface in (20) is defined as follows: $\Phi_E = c \oint_{S_k} F^{k0} \sqrt{-g}\, dS^{ij}$, where the indices $i, j, k = 1, 2, 3$ and they do not coincide with each other. In the limit of the special theory of relativity, the flux $\Phi_E$ turns into the flux of the electric field without additions from the magnetic field components.

From (19) we also obtain the integral theorem on the magnetic field circulation in the form of (22). In this case it is shown that if the thickness of the volume under consideration tends to zero, then (22) turns into integral equation (23), which generalizes the three-dimensional Kelvin-Stokes theorem.

Similarly, from four-dimensional covariant equation (32) we obtain the integral Gauss theorem for the magnetic field (33) and the integral Faraday's law of electromagnetic induction (35). In contrast to the three-dimensional approach, the covariant expression for the magnetic flux through the surface $S$ includes the determinant $g$ of the metric tensor: $\Phi = -\int_S g \mathbf{B} \cdot \mathbf{n} \, dS$.

The expression for the electromotive force also changes: $\varepsilon = -\oint_\ell g \mathbf{E} \cdot d\mathbf{r}$. According to (30), the electromotive force behaves as a potential with accuracy up to the multiplier $-g$: $\varepsilon = -g\varphi$.

Comparison of (29) and (33), (28) and (35) shows that the four-dimensional Kelvin-Stokes theorem (24) is presented within the framework of the special theory of relativity, and it should be replaced with the expression in the covariant notation (25). However, to ensure consistency of equations (28) and (35), it is necessary to assume that in (25) the determinant $g$ of the metric tensor does not depend on the time. In addition, equation (28) represents the law of electromagnetic induction only for a fixed contour with a constant area. It turns out that the four-dimensional Kelvin-Stokes theorem, even in the form of (25), does not allow us to fully describe the law of electromagnetic induction – this requires application of the divergence theorem to the dual electromagnetic field tensor, which leads to (35). Thus, the Kelvin-Stokes theorem turns out to be unnecessary in derivation of the integral equations of the electromagnetic field.

However, the advantage of the Kelvin-Stokes theorem is that equation (27) is obtained directly from it, in which the magnetic field flux $\Phi = -\int_S g \mathbf{B} \cdot \mathbf{n} \, dS$ through a certain fixed surface $S$ leads to circulation $\oint_\ell g \mathbf{A} \cdot d\mathbf{r}$ of the vector potential $\mathbf{A}$ around a fixed contour surrounding this surface. In addition, it becomes clear that changing of the magnetic field flux through the contour with the constant area leads to the vortex electric field in the contour due to change over time of the vector potential.

In addition to circulation, the vector potential $\mathbf{A}$ also has a flux according to integral equations (43-44). In this case, the flux of the potential $\mathbf{A}$ through a closed surface appears only when a change in time of a new physical quantity – the integral scalar potential $I_0 = \int_V A^0 \sqrt{-g} \, dx^1 dx^2 dx^3$ – occurs.

The obtained integral equations have been applied to describe the standard experiment with a frame, in which there is a moving crossbar, allowing us to change the area of the frame and the flux of the magnetic or electric field through the frame. As a rule, in such experiments, the signs of the fields' fluxes over the contour's area and of the fields' circulation around the contour are determined by the right-hand screw rule and the Lenz rule, respectively. If we proceed from the four-dimensional approach and equations (19) and (32), then the signs of the fluxes and circulation are determined automatically.

The analysis of the experiment on the electromagnetic induction with changing of the area in the magnetic flux shows incompleteness of the approach based on the Lorentz force in order to explain the appearance of the electromotive force. This approach turns out to be totally inapplicable in the case when the electric field flux changes as the area changes in the frame, which leads to the magnetic field circulation in the contour. The only approach, which allows us to explain the experiments with changing of the fluxes of both the magnetic and electric fields as the contour's area is changed, is recalculation of the electromagnetic field tensor components from the fixed reference frame $K$ into the reference frame $K'$ associated with the moving part of the contour (with the moving crossbar in the frame in Figure 1). At the same time, the circulation of the magnetic or electric field arising in $K'$, in $K$ manifests itself as the circulation of a certain effectively acting field, as manifestation of the corresponding effect in $K'$.

Thus, in the situation shown in Figure 1 initially there is no electric field and there is only a constant magnetic field. But as soon as the crossbar begins to move, the vortex electric field $E_b$ appears in it from the viewpoint of $K$, which leads to the circulation of this field. But in fact, in the crossbar, in its reference frame $K'$, the field $\mathbf{E}'_b$ emerges, as well as the corresponding circulation and the electromotive force $\varepsilon'_b = \gamma \varepsilon_b$, which in $K$ looks as $\varepsilon_b$. Hence, we can make the conclusion that the effects from the vortex electric and magnetic fields in the moving part of the contour from the viewpoint of $K$ are caused by the analogous effects in this part of the contour, when they occur from the viewpoint of the comoving reference frame $K'$.

If we proceed from (10), the magnetic field circulation in the contour can occur when the electric field flux changes over time due to changing of the value of the electric field itself. This is the consequence of application of the Kelvin-Stokes theorem to the fixed contour. However, the use of the four-dimensional approach, with the help of the divergence theorem, in view of (22-23) leads to a new effect in (42), according to which the magnetic field circulation in the contour can also occur when the electric field flux changes over time due to changing of the area bounded by the contour.